 \documentclass[smallabstract,smallcaptions]{dccpaper}

\usepackage{epsfig}
\usepackage{amsmath}
\usepackage{amssymb}
\usepackage{color}
\usepackage{url}
\usepackage{cite}
\usepackage{psfrag}
\usepackage{amsfonts,amssymb}
\usepackage{amsbsy}
\usepackage{graphicx}
\usepackage{array}
\usepackage{multirow}
\usepackage{bm}
\usepackage{subfigure}
\usepackage{verbatim}
\usepackage{colortbl}
\usepackage{stfloats}
\usepackage{bigstrut}
\usepackage{booktabs}
\usepackage{fancyhdr}
\usepackage[linesnumbered,ruled]{algorithm2e}
\usepackage{threeparttable}
\usepackage{dcolumn}
\usepackage{amsthm}

\newlength{\figurewidth}
\newlength{\smallfigurewidth}

\theoremstyle{plain}

\theoremstyle{definition}

\theoremstyle{Remark}

\newcommand{\st}{\textit{s}.\textit{t}., }
\newcommand{\ie}{\textit{i}.\textit{e}., }

\begin{document}

\title
{\large
\textbf{Linear Model based Geometry Coding for Lidar Acquired Point Clouds}
}

\author{%
\footnotesize Xiang Zhang, Wen Gao, and Shan Liu\\
\scriptsize Media Lab, Tencent America, Palo Alto, California, USA\\
\scriptsize \{xxiangzhang,~wengao,~shanl\}@tencent.com
}

\maketitle
\thispagestyle{empty}

\begin{abstract}
In this paper, we propose a new geometry coding method for point cloud compression (PCC), where the points can be fitted and represented by straight lines. The encoding of the linear model can be expressed by two parts, including the principle component along the line direction and the offsets from the line. Compact representation and high-efficiency coding methods are presented by encoding the parameters of linear model with appropriate quantization step-sizes (QS). To maximize the coding performance, encoder optimization techniques are employed to find the optimal trade-off between coding bits and errors, involving the Lagrangian multiplier method, where the rate-distortion behavior in terms of QS and multiplier is analyzed. We implement our method on top of the MPEG G-PCC reference software, and the results have shown that the proposed method is effective in coding point clouds with explicit line structures, such as the Lidar acquired data for autonomous driving. About 20\% coding gains can be achieved on lossy geometry coding.

\end{abstract}

\section{Introduction}
\label{sec:intro}
Nowadays, point clouds have becoming a new popular immersive multimedia and have been applied in a wide range of applications. In terms of the ultimate utility, the applications include human-oriented ones such as virtual reality (VR), augmented reality (AR) \cite{slfc_icip,slfc_jetcas}, and machine-involved ones such as autonomous driving and geographic information system. The human-oriented applications target on immersive and photo-realistic viewing experiences and therefore require relative high-resolution of point clouds. However, the machine-involved applications aim at high-precision in computer vision related tasks, where point clouds are generally acquired by Lidar sensors and the points are relatively sparser with weaker correlations across space.

A point cloud is generally a set of three-dimensional (3D) points, and each point is associated with extra attribute information including colors, reflectance, time stamps, etc.
The uniqueness of point clouds lies on the sparse and irregular distribution of the points in the space. Traditional image and video representations are pixels in regular 2D grids, where each position in the rectangular region is occupied and associated with a pixel value. However, for point clouds, only a small portion of the whole 3D space is occupied by physical points, while the rest of the space is empty.

With ever-increasing number of point cloud related applications, point cloud compression (PCC) becomes a critical issue, otherwise the storage and transmission of point clouds would be expensive and unrealistic. To resolve this, MPEG-3DG group has started the MPEG-PCC standard \cite{mpeg-pcc}, and it draws a lot of interests from both academic and industrial people and thus absorbs the most state-of-the-art PCC techniques.
In MPEG-PCC, there are two separate standards targeting at different application scenarios, \ie video-based PCC (or V-PCC) and geometry-based PCC (or G-PCC). V-PCC is essentially projecting 3D points onto 2D planes and then using existing video codec to compress the 2D planes, while G-PCC achieves compression in original 3D space by utilizing the 3D spatial and temporal correlations. For now, V-PCC is more efficient in coding dense point clouds and G-PCC is better for sparse point clouds in general. In this work, we focus on the geometry coding in G-PCC.

In G-PCC, the geometry of point clouds is coded by octree partition. The root node representing the whole 3D space is partitioned into 8 sub-nodes (sub-spaces), where the sub-nodes with occupied points can be further partitioned into 8 sub-nodes recursively until reaching the leaf nodes. To encode the octree structure, a `1' is coded if the sub-node is occupied and a `0' is coded otherwise. Thus for a non-leaf octree node, an 8-bit occupancy code is used to represent its 8 sub-node occupancy information. To further improve the coding efficiency, neighboring occupancy information of current sub-node is utilized as the context information in the bit-wise coding of the occupancy code.

\begin{figure}
\centering
\subfigure{\includegraphics[height=0.36\linewidth]{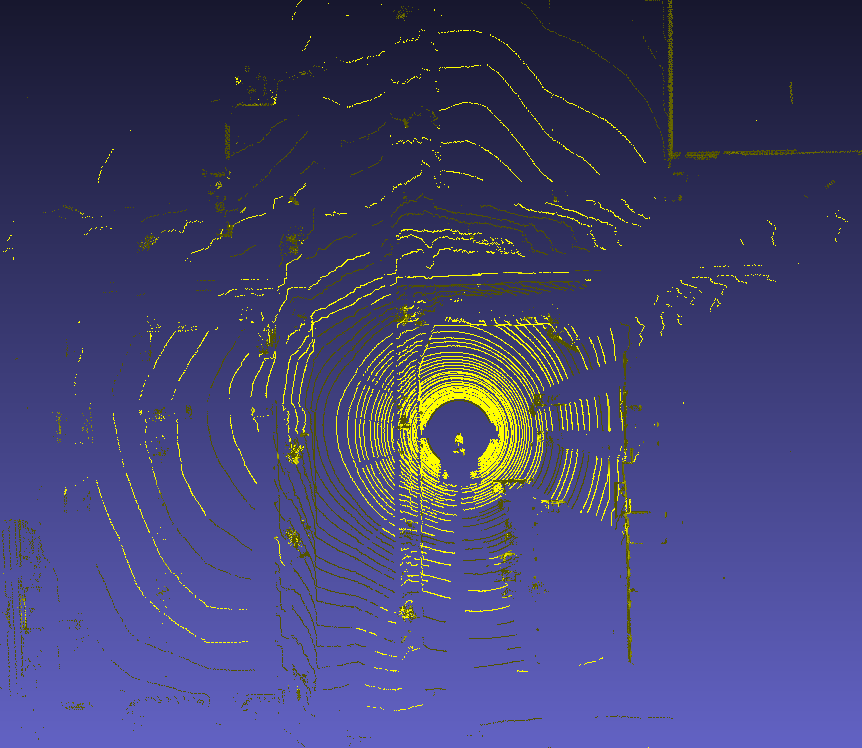}}
\subfigure{\includegraphics[height=0.36\linewidth]{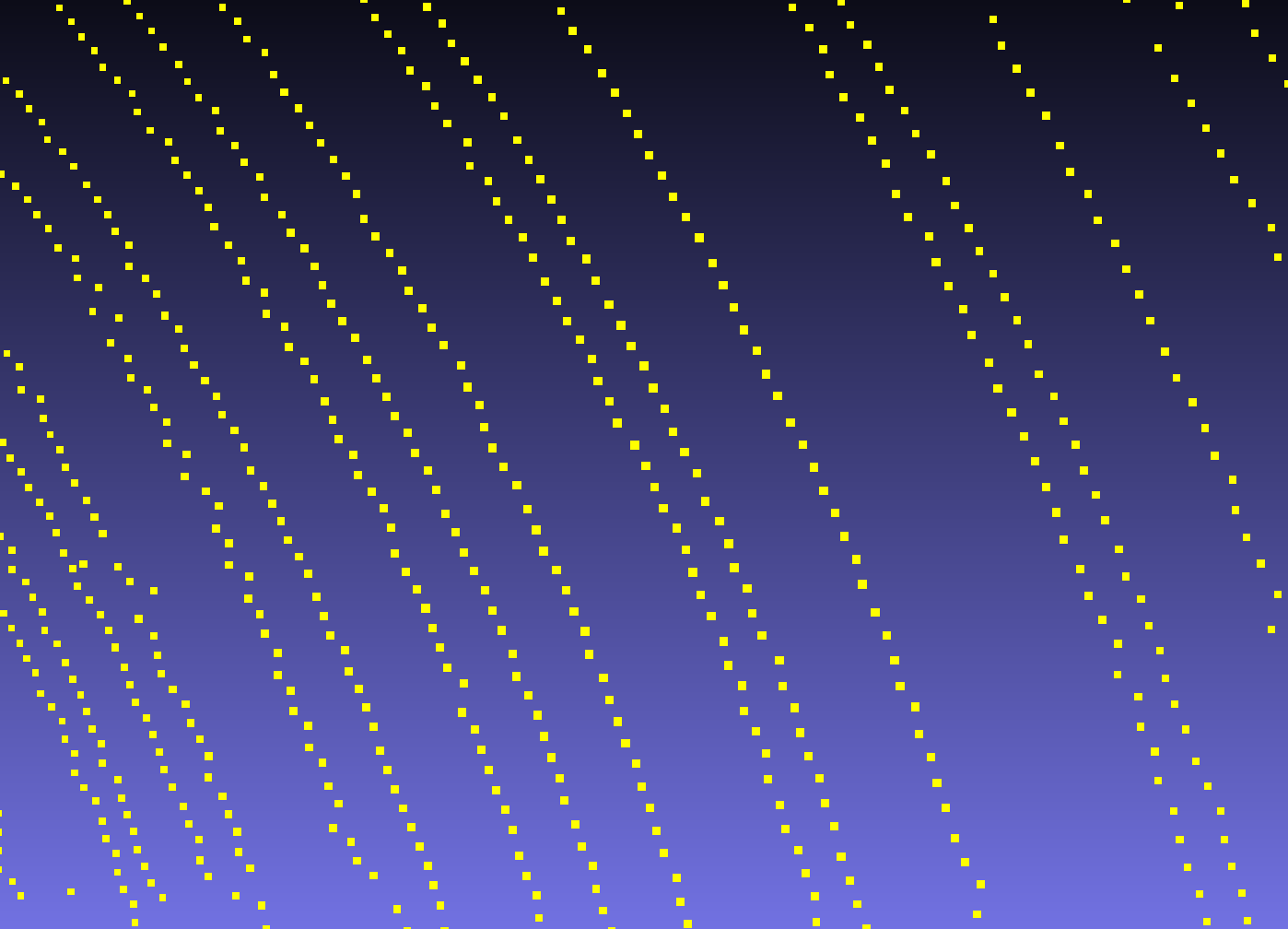}}
\caption{Zoom-out and zoom-in views of the ``Ford\_01\_q\_1mm-0100'' point cloud.}
\label{fig:example}
\end{figure}

However, the octree coding in current G-PCC has some limitations. First, it is not easy to utilize explicit geometry features in the octree structure especially for sparse point clouds, thus the compression efficiency may not be optimal. Second, octree coding introduces latency because of the hierarchical top-down tree structure and reference relationship to neighboring nodes.
A triangular-model based geometry coding method \cite{pavez2017dynamic} tries to fit triangular surfaces in 3D space and encodes the triangular parameters, which achieves good performance for dense point clouds at low bitrates and is adopted in G-PCC.
To demonstrate another particular geometry structure in Lidar obtained point clouds for autonomous driving, the ``Ford\_01\_q\_1mm'' sequence is a typical example. As shown in Figure~\ref{fig:example}, one can observe explicit line structures where points lie close to straight lines.

In light of above observations, in this work, we propose a new model-based method for geometry coding in G-PCC. More specifically, it is a linear model that takes the explicit line structures shown in Figure~\ref{fig:example} as a prior knowledge to improve the coding efficiency. The methods of line detection and refinement and the encoding of line parameters are introduced in this work. Moreover, the encoder-side techniques used to  optimize the rate-distortion performance are presented as well. The experimental results have shown that the proposed method is able to achieve significant gains on geometry coding for the Lidar acquired data, indicating the effectiveness and potential of the proposed linear model in the related application scenario.

The rest of the paper is organized as follows. In Section~\ref{sec:method}, we present the linear model and encoder optimization methods, followed by evaluation of the efficiency of the proposed method in Section~\ref{sec:experiment}. Finally, we conclude this paper in Section~\ref{sec:conclusion}.

\section{Linear Model based Geometry Coding for PCC}
\label{sec:method}

\begin{figure}
	\centering
	\includegraphics[width=1\linewidth]{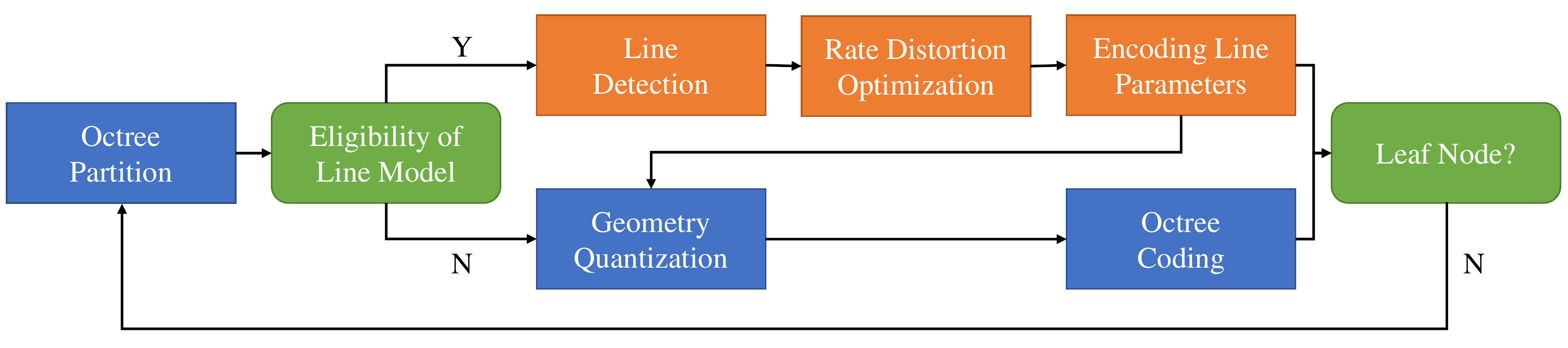}
	\caption{Framework of the linear model based geometry coding.}
	\label{fig:framework}
\end{figure}

The framework of the linear model based geometry coding is illustrated in Figure~\ref{fig:framework}. The proposed method is embedded into the octree coding in current G-PCC structure. For each sub-node in the octree, the eligibility of applying linear model is checked first. If it is eligible, the linear model is applied; otherwise, the same octree coding procedure as in current G-PCC is applied. The linear model starts from line detection procedure, in which the potential line candidates are detected from all the points in current sub-node. Subsequently, the line candidates are refined by optimizing the rate-distortion performance. Finally, the points close to the lines are encoded by line parameters. Note that the remaining points in current sub-node that cannot be efficiently represented by the linear model are then encoded by octree coding. The geometry coding ends when all the sub-nodes reach leaf nodes. The algorithm details will be elaborated in the following subsections.

\subsection{Line Detection}
The eligibility of the linear model is determined by point density, \ie the number of points in current node and the octree partition depth. Since this method targets at the sparse data obtained by Lidar systems, the point density criterion prevents applying linear model on dense point clouds where there is no obvious line structure. Furthermore, it would be inefficient and inaccurate to detect lines at the very beginning of the octree partitioning, because there will be extensive outliers and noises in line detection from a global point of view. Therefore, we only enable the line detection when the number of points in current node is within a reasonable range or the octree partition depth reaches a certain threshold. Figure~\ref{fig:example} is a good example, one would expect more efficient and accurate line detection from the local region on the right-hand side, instead of from the whole space on the left-hand side.

If the current node is eligible for linear model, lines are detected from the points in current node. Line detection is a mature algorithm in computer vision, any line detection methods can be applied here. In this work, we use the Iterative Hough Transform \cite{dalitz2017iterative} to detect lines in the 3D space. In a nutshell, it detects most possible lines in Hough space and uses Principle Component Analysis (PCA) to iteratively refine the lines.

Let $\bm X=\{\bm x_1, \bm x_2, \dots, \bm x_N\}$ be the collection of the points near a line detected by Hough transform, where $\bm x_i \in \mathbb R^{3\times1}$ denotes the Cartesian coordinates of a point in 3D space and $N$ is the number of points that are close to the line. The line can then be represented by two vectors $\bm a$ and $\bm b$, where $\bm a$ is a point right on the line and $\bm b$ is the angular vector which is a unit vector and parallel to the line. $\bm a$ can be initialized as the mass center of $\bm X$, \ie $\bm a=(\sum{\bm x_i})/N$. $\bm b$ is the eigenvector corresponding to the largest eigenvalue of $\bm Q^T \bm Q \in \mathbb R^{3\times3}$, where $\bm Q=(\bm x_i-\bm a)^T \in \mathbb R^{N\times3}$.

\subsection{Representation and Encoding of Lines}

\begin{figure}
\centering
\includegraphics[width=0.5\linewidth]{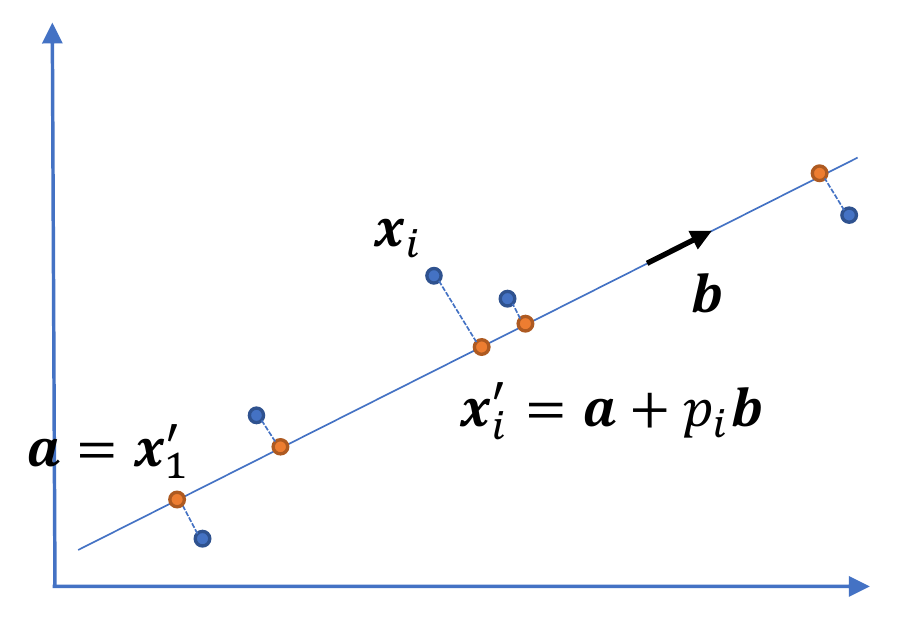}
\caption{Illustration of linear model in 2D. Blue dots are the original points, and red dots are their projections on the fitted line. The line can be represented by vectors $\bm a$ and $\bm b$, a point on the line and the direction of the line, respectively.}
\label{fig:line_fitting}
\end{figure}

In a nutshell, we represent the points close to a line by two parts, the first part is their projections on the line, which reflect the principle component of these points; and the second part is the offsets from the line. This representation is more compact because of the fact that the principle component can be coded more efficiently and can recover the positions with high accuracy.

The line representation is shown in Figure~\ref{fig:line_fitting}. Each point $\bm x_i$ can be projected to the line by $\bm x'_i=\bm P\bm x_i+\bm a-\bm P\bm a$, where $\bm x'_i$ is the projected point and $\bm P=(\bm b^T \bm b)^{-1} \bm b\bm b^T$ is the corresponding projection matrix, which can be further simplified as $\bm P=\bm b\bm b^T$, as $\bm b$ is unit. We then reorder these points by their projection position on the line relative to point $\bm a$, \ie for any pair of $\bm x_i$ and $\bm x_j (i<j)$, they satisfy that $p_i<p_j$, where $p_i=\bm b^T(\bm x'_i-\bm a)=\bm b^T(\bm x_i-\bm a)$ is the projection position of $\bm x_i$ on the line relative to point $\bm a$. Without loss of generality, one can let $\bm a=\bm x'_1$ such that $p_1=0$ and $p_i>0$ for $i>1$.

To be efficient, we propose to represent and encode the points close to a line by three parts, including \textit{line parameters}, \textit{projection parameters} and \textit{offset parameters}.

First of all, \textit{line parameters} include the starting point $\bm a$ and the angular vector $\bm b$. Encoding the starting point $\bm a=(a_x,a_y,a_z)^T$ is straightforward by its Cartesian coordinates relative to the origin of current octree node. The angular vector $\bm b=(b_x,b_y,b_z)^T$ is proposed to encode in spherical coordinates, \ie $(\theta,\phi)^T$, where $\theta=\arctan(\frac{b_y}{b_x})$ and $\phi=\arctan(\frac{b_z}{\sqrt{b_x^2+b_y^2}})$. The benefit of introducing spherical coordinates is not only reducing one redundant dimension (with the prior of $\bm b$ is unit), but also enabling more efficient context modeling in entropy coding. As shown on the right-hand side in Figure~\ref{fig:example}, one can observe that the lines detected from a local region are appeared to be parallel to each other, indicating they may have similar values in $(\theta,\phi)$ across space and thus it would be more efficient for compression.
Moreover, redundancies can be further reduced by considering the symmetric property of $\bm b$, this can be achieved by restricting $\phi \ge 0$ and restricting $\theta \ge 0$ when $\phi=0$.
Note that the number of points on each line $N$ should be specified in the bitstream as well.

\begin{figure}
\centering
\includegraphics[width=0.65\linewidth]{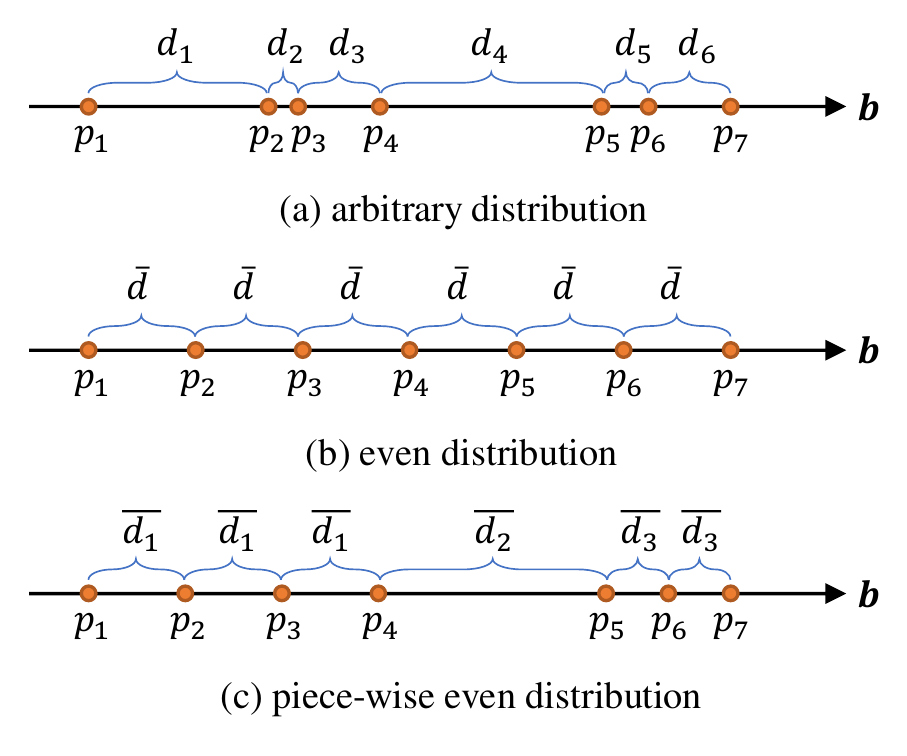}
\caption{1D illustration of three encoding modes of projection parameters.}
\label{fig:line_projection}
\end{figure}

Next, \textit{projection parameters} are a sequence of $\{p_i\}$ for $i=1,\dots,N$. Because of the reordering process, $p_i$ is in ascending order. Instead of encoding the value of $p_i$ directly, we propose to encode their differences, \ie $\{d_i\}$, where $d_i=p_{i+1}-p_i$ and $p_1$ is always 0 because $\bm a=\bm x'_1$.
Note that the projection from the original point $\bm x_i$ to the line $\bm x'_i$ does not have to be orthogonal for the purpose of efficient compression.
Figure~\ref{fig:line_projection} shows three projection modes in 1D, where 7 points are projected onto a line. Arbitrary mode in Figure~\ref{fig:line_projection}(a) encodes the differences of two neighboring projections directly, \ie $d_1,\dots,d_6$.
However, this mode would be expensive at lower bitrate. An alternative option would be encoding an average value of $d_i$, \ie $\bar{d}=\frac{\sum{d_i}}{N}$, by assuming all the projection positions are evenly distributed along the line as shown in Figure~\ref{fig:line_projection}(b).
This may inevitably decrease the approximation accuracy, but it is a better trade-off of saving much more coding bits at low bitrate. The third possible mode is a compromise between the two aforementioned modes, which is the piece-wise even mode, where projection positions are assumed to be piece-wise evenly distributed along the line as shown in Figure~\ref{fig:line_projection}(c).

Finally, \textit{offset parameters} are the residual vectors from the projection position $\bm x'_i$ to the original position $\bm x_i$, \ie $\bm r_i=\bm x_i-\bm x'_i$. Lossless coding can be achieved if offsets are compressed losslessly.
From the compression point of view, encoding \textit{line parameters} plus \textit{projection parameters} resembles predictive coding while encoding \textit{offset parameters} are residual-like coding.

\subsection{Quantization and Reconstruction}

Note that all the parameters can be quantized before encoding for different bitrate conditions. In this work, we mainly focus on lossy geometry coding. Therefore, we apply the even-distribution mode as shown in Figure~\ref{fig:line_projection}(b) and ignore the \textit{offset parameters}, indicating the points are assumed to be evenly located right on the detected line. In other words, we only encode the principle component of the points along the line.

The quantization of the angular vector $\bm b=(\theta,\phi)$ is performed by dividing the sphere space. Since $\theta \in [-\pi,\pi]$ and $\phi \in [0, \pi/2]$, we divide $\phi$ into $Q_a$ equal segments and divide $\theta$ into $Q_a\times4$ equal segments to make sure the quantization in two directions has the same resolution. In our experiments, we set $Q_a=40$, which gives the best trade-off between coding overheads and approximation accuracy.

The quantization of the line starting point $\bm a=(a_x,a_y,a_z)$ and the averaged projection difference $\bar d$ are relevant, as they directly reflect the expected geometry accuracy. Suppose the  quantization step-size (QS) of geometry coding is $Q_g$, we then set the QS of $\bm a$ to be $Q_s=Q_g$ and the QS of $\bar d$ to be $Q_d=\frac{Q_g}{N-1}$. Note that $Q_d$ equals to $Q_g$ divided by $N-1$, because the errors caused by $\bar d$ would be accumulating.

The quantized parameters are then coded by entropy coder. Let $\widehat{\bm a}$, $\widehat{\bm b}$ and $\widehat{d}$ be the dequantized parameters of $\bm a$, $\bm b$ and $\bar d$, respectively. The reconstructed geometry position of point $\bm x_i$ on the line can be simply done by $\widehat{\bm x}_i=\widehat{\bm a}+\widehat{p}_i \widehat{\bm b}$, where $\widehat{p}_i=(i-1) \cdot \widehat{d}$ is the reconstructed projection parameter.
Note that if the offset parameters are encoded as well, the reconstruction would be $\widehat{\bm x}_i=\widehat{\bm a}+\widehat{p}_i \widehat{\bm b}+\widehat{\bm r}_i$, where $\widehat{\bm r}_i$ indicates the reconstructed offsets after dequantization.

\subsection{Rate-Distortion Optimization}

A proper rate-distortion optimization (RDO) is indispensable to adaptively determine a ``best'' mode in terms of data characteristics and bitrate conditions. 
In this subsection, we propose a fast RDO scheme, in which a subset of points is selected out of the points that are initially detected as on a line, such that the optimal rate and distortion trade-off can be approximated.

RDO theory states that the best mode minimizes the distortion under a rate constraint, or equivalently the best mode minimizes the rate with a distortion constraint. In the context of PCC, the distortion and rate terms actually indicate the average distortion and average rate over the total number of encoded points, which can be formulated as follows,
\begin{equation}
\min{\frac{R}{N}}, \quad \st \frac{D}{N}<\bar{D_c},
\label{eq:rdo}
\end{equation}
where $R$ and $D$ are the overall rate and distortion, respectively, and $N$ is the number of points, and $\bar{D_c}$ denotes the maximum tolerant distortion on average.

Suppose we have a group of points $\bm X=\{\bm x_1,\bm x_2,\dots,\bm x_N\}$ that is close to a detected line, and denote their projection positions relative to $\bm a$ as $\{p_i\}$ and the distance between two adjacent projection positions as $d_i=p_{i+1}-p_i$ for $(i=1,2,\dots,N-1)$. The problem then is to find an optimal subset $\bm X_{(i,j)}=\{\bm x_i,\dots,\bm x_j\}$ from $\bm X$ that minimizes the following equation,
\begin{equation}
\min_{1\le i<j\le N}{\bar R_{(i,j)}}, \quad \st \bar D_{(i,j)}<\bar{D_c},
\label{eq:rdo2}
\end{equation}
where $\bar R_{(i,j)}$ and $\bar D_{(i,j)}$ represent the average bitrate and average distortion when encoding the subset $\bm X_{(i,j)}$, respectively. However, it would be rather costly to perform a full RDO for every subset, because the number of all possible subsets is at least $O(N^2)$. Therefore, we propose the following fast algorithm by simplifying the rate-distortion model in \eqref{eq:rdo2}.

First, the rate model can be simplified as $\bar R_{(i,j)}=\frac{R}{j-i+1}$, where the total rate $R$ can be viewed as a constant, because only the averaged projection parameter $\bar d$ is encoded regardless of which subset is opted. Therefore, the longer the subset is, the less averaged bits can be achieved.
Second, the distortion model can be estimated by the $\ell_1$ norm as follows,
\begin{equation}
\bar D_{(i,j)}=\frac{\|(e_i,e_i+e_{i+1},\dots,\sum_{k=i}^{j-1}{e_k})\|_1}{j-i},
\label{eq:distortion_model}
\end{equation}
where $e_t=d_t-\frac{\sum_{k=i}^{j-1}{d_k}}{j-i}$ is the error between $d_t$ and the averaged projection distance of $\bm X_{(i,j)}$.
By substituting $\bar R_{(i,j)}$ and $\bar D_{(i,j)}$ into \eqref{eq:rdo2}, the RDO model can be simplified to finding a longest subset whose average distortion is below the maximum tolerant distortion $\bar D_c$ as follows,
\begin{equation}
\max_{1\le i<j\le N}{(j-i)}, \quad \st \bar D_{(i,j)}<\bar D_c,
\label{eq:rdo3}
\end{equation}
which can be solved much more efficiently.

After determining the optimal subset by the fast RDO, a full RDO can be performed to calculate the actual rate-distortion score (RDS) by linear model as follows,
\begin{equation}
RDS_l = P_l(Q_g)-\lambda R_l(Q_g),
\label{eq:rdo_real}
\end{equation}
where $P_l$ and $R_l$ are the actual geometry PSNR and the actual coding bits by linear model, respectively, which are functions of the geometry QS, \ie $Q_g$. Note that we use the geometry PSNR instead of geometry distortion because the range of geometry distortion may vary dramatically for different scenes and geometry PSNR is a normalized distortion in terms of the maximum geometry value. $\lambda$ is the well-known Lagrangian multiplier. The larger $RDS_l$ indicates a better rate-distortion performance under a certain value of $\lambda$.
If the RDS of linear model is larger than a threshold, \ie $RDS_l > T$, it indicates that linear model has a better rate-distortion trade-off than octree coding, and linear model will be applied for current node in this case, as $T$ is a threshold that reflects the RDS by octree coding. Otherwise, the octree coding is applied.
The training of optimal $Q_g$ and $T$ in terms of $\lambda$ is elaborated in Section~\ref{sec:experiment_rdo}.

\section{Experimental Results}
\label{sec:experiment}
We implement the linear model on top of the MPEG G-PCC reference software, \ie TMC13v6 \cite{tmc13}.
Three Lidar acquired point clouds, ``Ford\_01\_q\_1mm'', ``Ford\_02\_q\_1mm'' and ``Ford\_03\_q\_1mm'' are utilized to demonstrate the effectiveness of the proposed method. One frame of ``Ford\_01\_q\_1mm'' is shown in Figure~\ref{fig:example}.

\subsection{Rate-Distortion Optimization}
\label{sec:experiment_rdo}
To derive the optimal rate-distortion behavior, a training process is involved to obtain the optimal geometry QS, \ie $Q_g$, and the threshold $T$ in terms of $\lambda$. The training resembles that in video coding and feature coding \cite{rdo_video,joint_feature}.
First, for each fixed $\lambda$, we sweep possible combinations of $Q_g$ and $T$, and then find the optimal one that minimizes the rate-distortion cost in \eqref{eq:rdo_real}.
Then, we vary $\lambda$ from 0 to 30 by a step of 5 and optimize $Q_g$ and $T$ accordingly.

\begin{figure}
\centering
\subfigure{\includegraphics[width=0.48\linewidth]{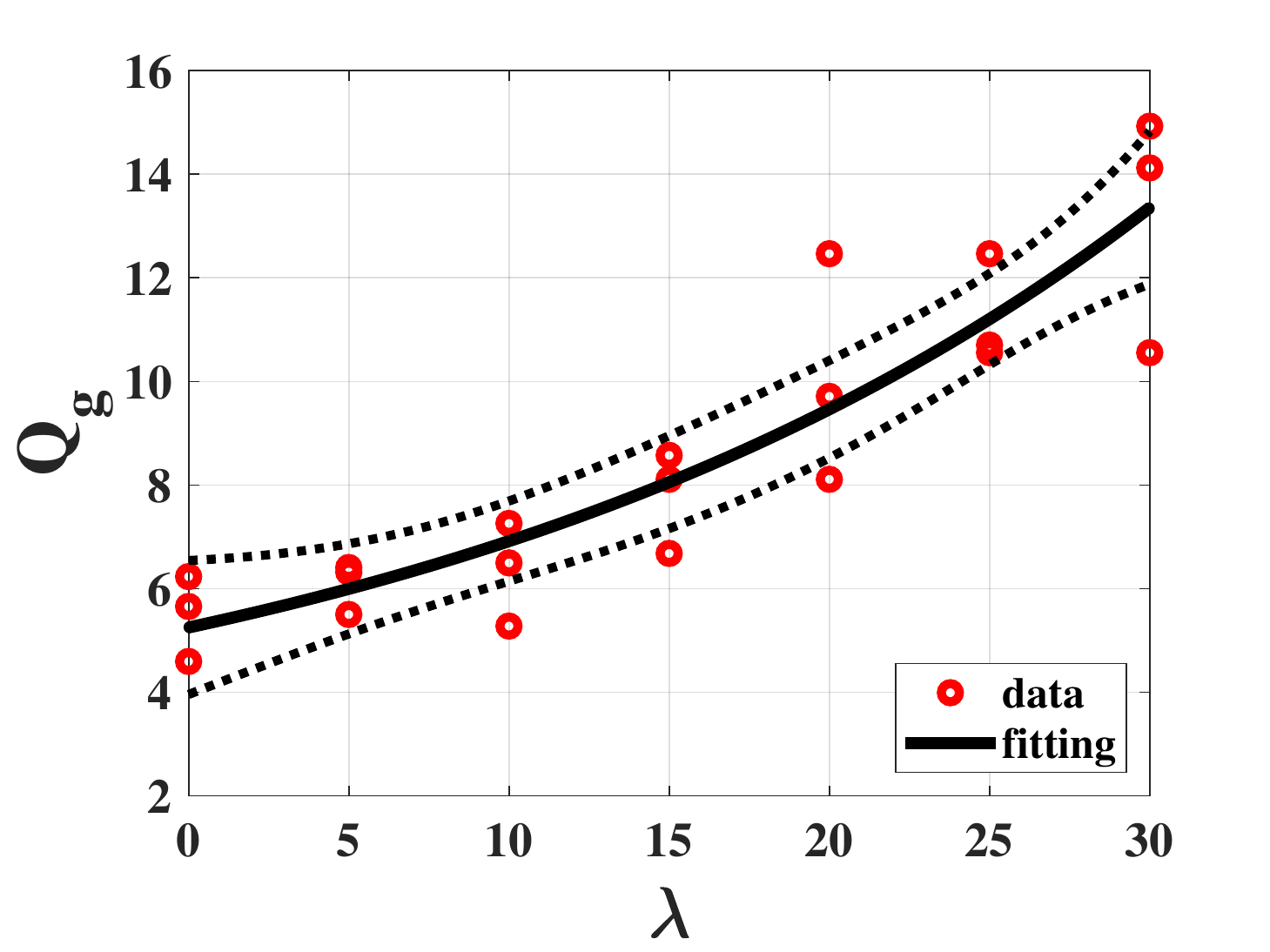}}
\subfigure{\includegraphics[width=0.48\linewidth]{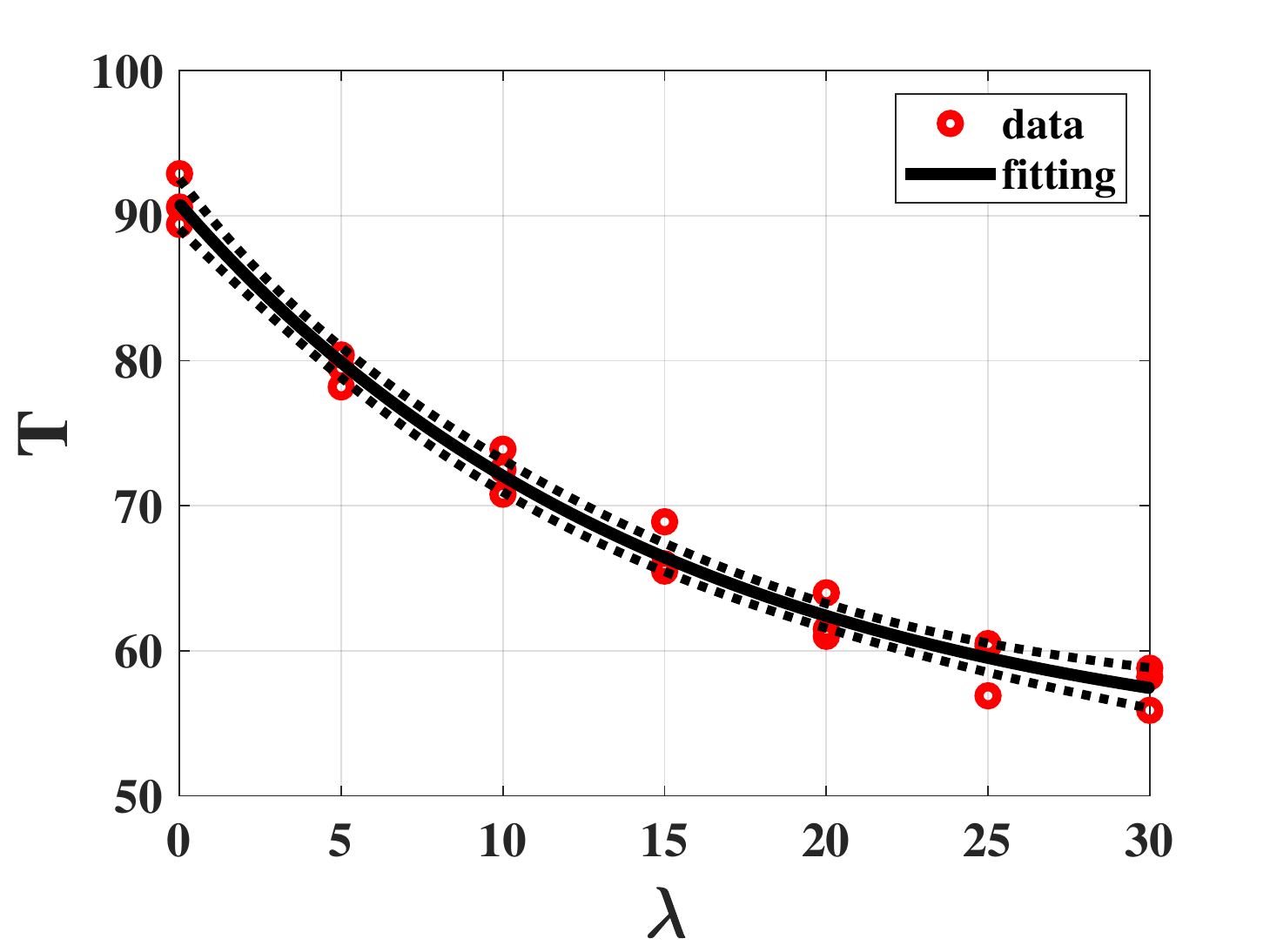}}
\caption{Fitting optimal $Q_g$ and $T$ as functions of $\lambda$ with 95\% confidence bounds.}
\label{fig:fit_lambda_qp_T}
\end{figure}

The relationship between optimal $Q_g$ and $\lambda$ is fitted by an exponential function, and the relationship between optimal $T$ and $\lambda$ is fitted by a reciprocal exponential function, as shown in Figure~\ref{fig:fit_lambda_qp_T}. From the figure, one can observe that with the increasing of $\lambda$ the optimal $Q_g$ increases and the optimal $T$ decreases monotonously. It makes sense because a larger $\lambda$ indicates the situation that the bitrate is more important than distortion, in which a larger $Q_g$ is applied for achieving lower bitrate and a smaller $T$ estimates a lower RDS of octree coding at low bitrate conditions.

\subsection{Experimental Results}
\label{sec:experiment_result}

The lossy geometry coding under C2 condition \cite{gpcc_ctc} was tested, where two extra operation points are added to make the bitrate increases more smoothly. The quality of the reconstructed geometry is evaluated by D1 and D2 metrics. D1 measures the point-to-point distortions while D2 measures the point-to-plane distortions \cite{tian2017geometric}.

\begin{figure}
\centering
\subfigure[ford\_01\_q1mm]{\includegraphics[width=0.48\linewidth]{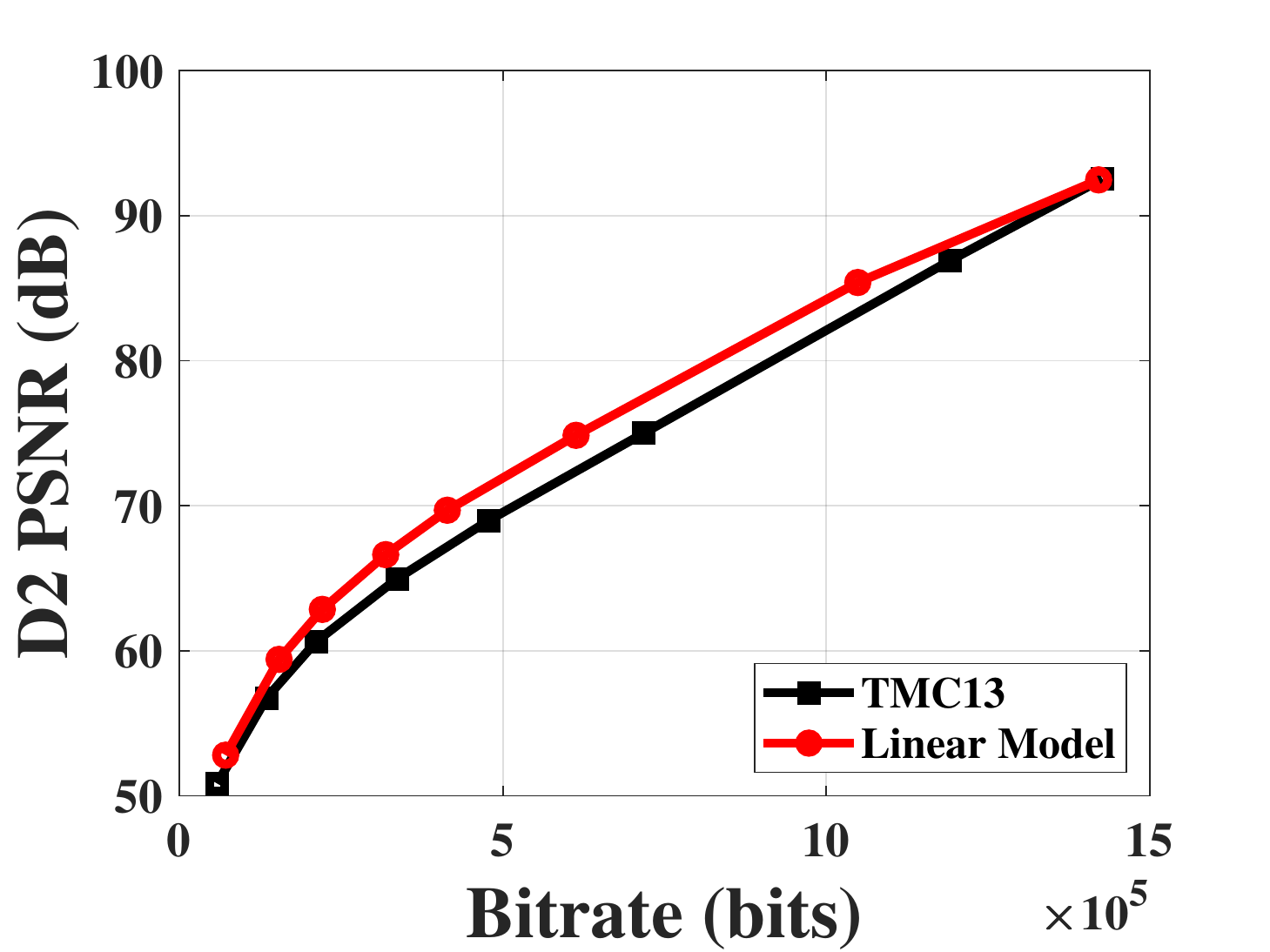}}
\subfigure[ford\_03\_q1mm]{\includegraphics[width=0.48\linewidth]{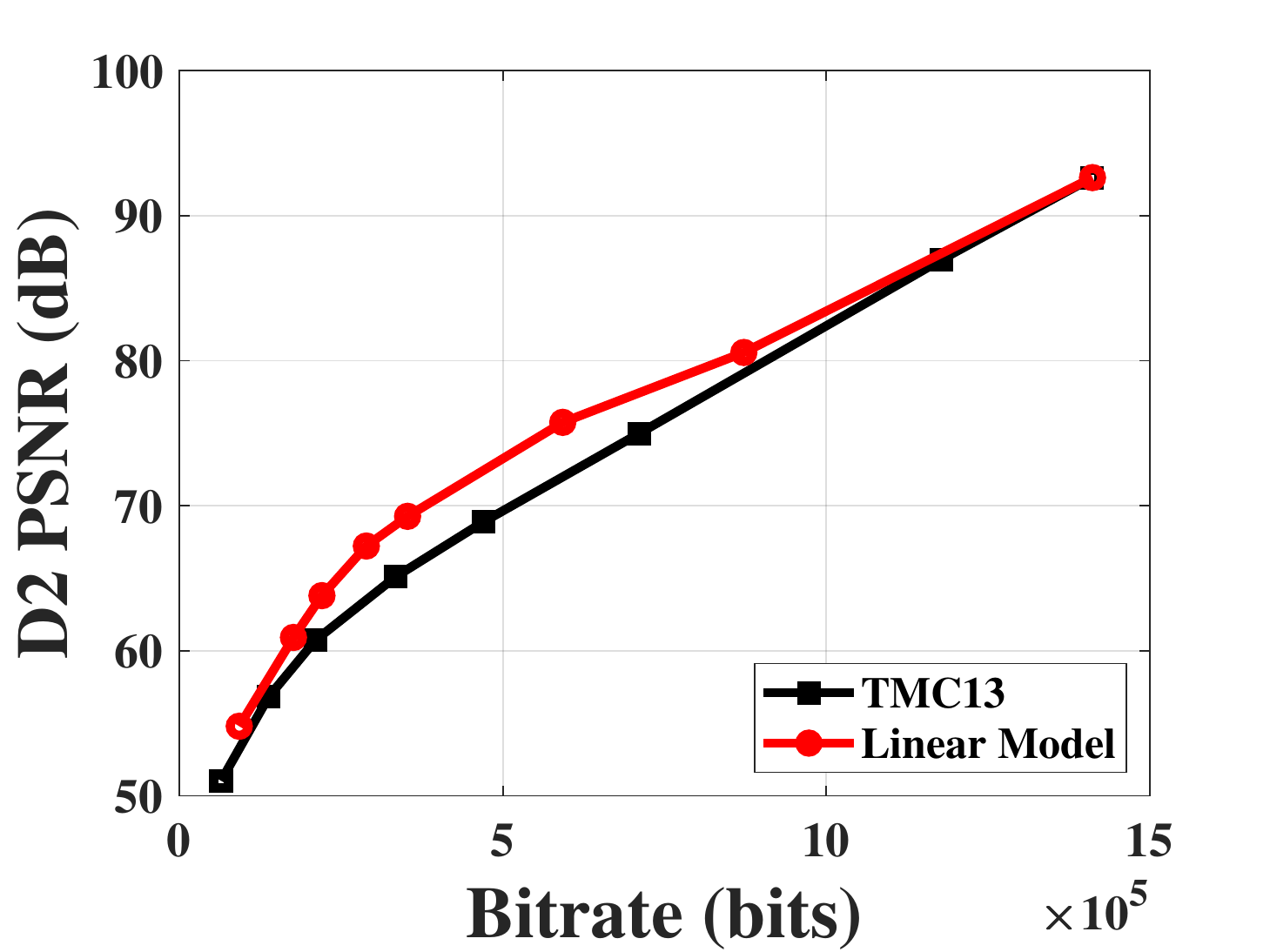}}
\caption{Rate-distortion curves in terms of bitrate and D2 PSNR.}
\label{fig:rdcurve}
\end{figure}

We compare the linear model with the TMC13 anchor by BD-rate \cite{bdrate}. As shown in Table~\ref{tab:bdrate}, the linear model can achieve 4.4\% coding gains in terms of D1 and 19.8\% coding gains in terms of D2, on average, for the Lidar acquired data. In the context of machine-oriented application scenarios, it is acknowledged that D2 is a more important metrics because it measures the structure similarity in 3D and those structural features are critical in computer vision tasks. The results indicate that the proposed linear model is able to preserve more geometry structures than octree coding at the same bitrate, because compressing the points on a line along the principle component is more efficient. Two rate-distortion curves are shown in Figure~\ref{fig:rdcurve}, where one can see that the linear model shows more gains at middle and lower bitrates.

\begin{table}
  \centering
  \caption{BD-rate against TMC13 of lossy geometry coding.}
    \begin{tabular}{|l|cc|}
    \hline
    \multicolumn{1}{|c|}{\textbf{C2}} & \multicolumn{2}{c|}{\textbf{Geometry BD-rate}} \bigstrut[t]\\
    \multicolumn{1}{|c|}{\textbf{Sequences}} & \textbf{D1} & \textbf{D2} \bigstrut[b]\\
    \hline
    \textbf{ford\_01\_q1mm} & -5.7\% & -16.5\% \bigstrut[t]\\
    \textbf{ford\_02\_q1mm} & -4.4\% & -20.5\% \\
    \textbf{ford\_03\_q1mm} & -3.2\% & -22.2\% \bigstrut[b]\\
    \hline
    \textbf{Average} & \textbf{-4.4\%} & \textbf{-19.8\%} \bigstrut\\
    \hline
    \end{tabular}%
  \label{tab:bdrate}%
\end{table}%

\section{Conclusion}
\label{sec:conclusion}
In this paper, we present a model-based geometry coding method for Lidar acquired point clouds, where the linear model fits the points to straight lines. The linear model can be efficiently encoded by the principle component of the line. Rate-distortion optimization (RDO) techniques are introduced to improve the overall coding performance. We implement the linear model on top of the MPEG G-PCC reference software, and the experimental results have shown significant coding gains for Lidar acquired data.

%\footnotesize{
\Section{References}
\bibliographystyle{IEEEtran}
\bibliography{IEEEabrv,strings}
%}

\end{document}